# Born Oppenheimer Dynamics Near Metal Surfaces

Michael Galperin[1], Abraham Nitzan[2] and Mark Ratner[3]

[1] Department of Chemistry and Biochemistry, University of California at San Diego, La Jolla, CA 92093, USA
[2] School of Chemistry, The Sackler Faculty of Science, Tel Aviv University, Tel Aviv 69978, Israel
[3] Department of Chemistry and Materials Research Center, Northwestern University, Evanston, IL 60208, USA

Electron exchange processes between molecules adsorbed on metal surfaces and the underlying metal are common in interfacial chemistry, encompassing nearly all electrochemical phenomena and molecular electronics and playing important roles in many other molecular processes at metal surfaces. In many processes, including electrochemical electron transfer and molecular conduction at metal-molecule-metal junctions, a central issue is the effect of nuclear motions on the characteristics of the electronic process. In many others, including electron induced desorption, electron tunneling induced reactions and friction involving molecules at metal surface, the focus is on the nuclear dynamics itself. Understanding the nature of this dynamics is a potentially complex issue, since a fundamental cornerstone of any molecular theory, the Born Oppenheimer (BO) principle which underlines the meaningful existence of nuclear potential surfaces, is called here into question. Indeed, while in most molecular processes timescale separation between nuclear and electronic motion makes it possible to define meaningful nuclear potential surfaces and to discuss non-adiabatic transitions between them as resulting from relatively rare instances of breakdown of this separation, the dense electronic spectrum of metals and the slow electronic timescale $\hbar\rho$ associated with the electronic density of states $\rho$, renders the BO approximation questionable in such systems.

In dealing with this problem, it is useful to consider limiting cases. If we represent the metal as a free-electron reservoir and the metal-molecule interaction by the usual electron transfer Hamiltonian $\sum_{k,j}\left(V_{jk}\hat{d}_j^\dagger \hat{c}_k + V_{kj}\hat{c}_k^\dagger \hat{d}_j^\dagger\right)$ where $\hat{d}_j$ and $\hat{c}_k$ are single electron annihilation operators on the molecular electronic orbital $j$ and metal level $k$, respectively, the timescale for electron exchange between the molecule and the metal is given by $\Gamma = 2\pi V^2 \rho$, evaluated at the molecular level energy $E_j$. When $\Gamma \to 0$ we should recover the molecular BO picture, and for small finite $\Gamma$ electron transfer events mark the non-adiabatic crossover (surface hopping) between different molecular potential surfaces that characterize the different redox states of the molecular species.[1] These potential surfaces may be somewhat modified by the proximity of the metal, e.g. by interaction of the molecule with the image of its charge distribution in the metal, however it is safe to assume that the usual electronic-nuclear timescale separation is not affected by this interaction, i.e. that the metal electronic response is fast relative to the molecular nuclear timescale.

The rest of this note concerns the other limit $\Gamma \to \infty$, or more precisely $\Gamma \gg \omega$ where ω is a characteristic vibrational frequency (i.e. $\hbar\omega^{-1}$ is the characteristic nuclear timescale). Denote the electronic and nuclear subsystem of the molecule $S$ by $A$ and $B$, respectively so that the Hamiltonian of the isolated molecule is $\hat{H}_S = \hat{H}_A + \hat{H}_B + \hat{V}_{AB}$. Let subsystem $A$ be characterized by two states, $|1_A\rangle$ and $|2_A\rangle$ (e.g. two charging states of the molecule, in which case B represents the relevant underlying nuclear motion). Assume furthermore that an additional process (e.g. electron exchange with a nearby metal) causes a rapid interchange between states $|1_A\rangle$ and $|2_A\rangle$, and that this process by itself would bring subsystem $A$ to equilibrium characterized by a density matrix $\rho_A$. A Born-Oppenheimer-type approximation suggests that in the limit where the interchange dynamics between $A$ states is fast relative to the characteristic dynamics of B, the motion of B will be described to a good approximation by the mean field Hamiltonian $H_B + \langle V_{AB} \rangle_A$ where $\langle ... \rangle_A = \text{Tr}(\rho_A ...)$.

When $A$ represents the electronic states of an adsorbed molecule and $B$ – its nuclear subspace, the resulting nuclear potential surface will often bear little resemblance to that of the isolated molecule. When the molecular nuclear dynamics is represented by a single harmonic oscillator coupled to the electronic subsystem by the standard polaron coupling model (charge dependent shift of the harmonic potential surface) we have shown[2] that the resulting potential surface can possess more than one minimum, implying a potential multi-stable behavior. This result was obtained using the mean field (MF) argument above, and we have recently[3] re-derived it by evaluating the electronic Green function G in the coupled system using the linked cluster approximation. This procedure yields the standard isolated molecule result in the isolated molecule limit, $\Gamma \to 0$, as well as the MF result in the static limit, $\omega_0/\Gamma \to 0$.

The validity of latter derivation was recently questioned by Alexandrov and Bratkovsky.[4] In particular, they question our use of the non-equilibrium linked cluster expansion (NLCE), suggesting that using the renormalized level occupation instead of the bare one in the exponentiated cluster amounts to overcounting diagrams in the perturbation expansion of the electronic Green function. We believe that this claim is wrong. To make the consideration simple let us focus on the relevant static limit, $\omega/\Gamma \to 0$, and disregard all the diagrams with non-zero phonon frequency,[5] this leaves in consideration diagrams of the Hartree type, $\Sigma_H$, only (see Fig. 1a). The standard linked cluster expression sums the corresponding clusters to all orders, leading (after expansion of exponent into series) to a Dyson-type equation for the Green function

$$G \approx G_0 e^{\Sigma_H G_0} \to G_0 + G_0 \Sigma_H G_0 + G_0 \Sigma_H G_0 \Sigma_H G_0 + \ldots \qquad (1)$$

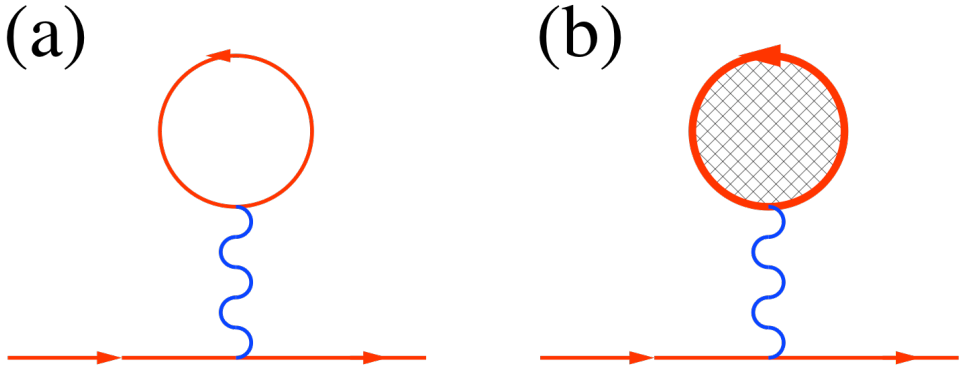

Fig. 1. The Hartree diagram $G_0 \Sigma_H^{(2)} G_0$ (a) and the dressed Hartree diagram $G_0 \Sigma_H G_0$ (b) used in the linked cluster expansion of the electronic self-energy

Instead, in Ref.[3] we have replaced the bare $\Sigma_H^{(2)}$ ♣

---

♣ In Eq. (2), $M$ is the electron-phonon coupling and all functions are defined on the Keldysh contour c. $G_0$ and $D_0$ are free electron and free phonon Green functions, respectively, and $G_0(\tau_1, \tau_1+)$ corresponds to the limit $\lim_{\tau_2 \to \tau_1+} G_0(\tau_1, \tau_2)$.

$$\Sigma_H^{(2)}(\tau,\tau') = \delta(\tau,\tau') M^2 \int_c d\tau_1 D_0(\tau,\tau_1) \left[-i G_0(\tau_1,\tau_1+)\right] \qquad (2)$$

by the renormalized self-energy $\Sigma_H$ in which the $G_0$ in (2) (that corresponds to the bare (blank) bubble in Fig. 1a) is replaced by the exact Green function $G$ (corresponding to the dressed (grey) bubble in Fig. 1b). Fig. 2 shows some of the terms that constitute the dressed bubble.

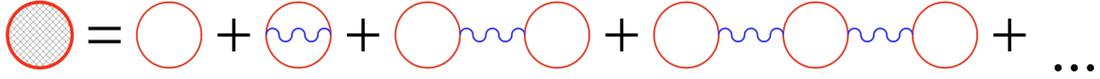

Fig. 2. Bubble diagrams contributing to the renormalized self-energy $\Sigma_H$

The authors of Ref [4] have questioned this renormalization procedure, claiming that it leads to double counting in the perturbation expansion of the electronic Green function in the electron-vibration coupling. We disagree. In fact the diagrams (Fig. 2) that contribute to the renormalization leading to diagram (Fig. 1b) are topologically different than those used in the standard linked cluster expansion in diagram (1a), so the latter can not account for the dressing processes that results in diagram (1b).

We further demonstrate this point by evaluating the self-energy to 4$^{th}$ order in electron-vibration coupling. As pointed out above, in the static limit only diagrams of the Hartree type are of interest. Contributions of this type that corresponds to the third and second diagrams in Fig.2 are

$$\Sigma_H^{(4H)}(\tau,\tau') = M^2 \delta(\tau,\tau') \int_c d\tau_1 D_0(\tau,\tau_1) \left[-i \int_c d\tau_2 \int_c d\tau_3 G_0(\tau_1,\tau_2) \Sigma_H^{(2)}(\tau_2,\tau_3) G_0(\tau_3,\tau_1+)\right]$$

$$\Sigma_H^{(4X)}(\tau,\tau') = M^2 \delta(\tau,\tau') \int_c d\tau_1 D_0(\tau,\tau_1) \left[-i \int_c d\tau_2 \int_c d\tau_3 G_0(\tau_1,\tau_2) \Sigma_X^{(2)}(\tau_2,\tau_3) G_0(\tau_3,\tau_1+)\right]$$

respectively. Here $\Sigma_X^{(2)}(\tau,\tau') = i M^2 D_0(\tau,\tau') G_0(\tau,\tau')$ is the Born (Fock, exchange) self-energy. It is easy to show that these contributions are absent in the case of an isolated molecule. They do exist however for a molecule in a junction.

In summary, Born-Oppenheimer type potential energy surfaces can be defined and used for molecules strongly coupled ($\Gamma \gg \omega_0$) to metal surfaces. Using the undressed cluster in the linked cluster expansion, as proposed in Ref. [4], misses the feedback effect: not only do tunneling electrons shift the position of the vibration (physics contained in diagrams of type a in Fig. 1), but also shifted (polarized) vibration affects tunneling electrons (dressing presented in type b diagrams). We believe that both effects should be present in a complete theory.